\documentclass[a4paper,twocolumn,color]{esapub} 

\newcommand{\targ}{IGR~J19140+098}
\newcommand{\GRS}{GRS~1915+105}
\newcommand{\inte}{INTEGRAL}

\usepackage{natbib}
\usepackage{psfig}
\usepackage{graphicx}
\title{\targ: a new \inte~transient}
\author[1]{Juho Schultz (juho.schultz@astro.helsinki.fi)}
\author[1]{Diana C. Hannikainen}
\author[1]{Osmi Vilhu}
\author[2]{Jerome Rodriguez}
\author[3]{Clement Cabanac}
\author[3]{Gilles Henri}
\author[3]{Pierre-Olivier Petrucci}
\author[1]{Panu Muhli}
\affil[1]{Observatory, University of Helsinki, 00014 University of
Helsinki, Finland}
\affil[2]{CEA Saclay, 91191 Gif-sur-Yvette, FRANCE \& ISDC,
Chemin d'Ecogia 16 1290 Versoix, Switzerland}
\affil[3]{Laboratoire d'Astrophysique, Observatoire de Grenoble,
38041 Grenoble Cedex 9, France}

\begin{document}


\maketitle

\begin{abstract}
\targ~is a new X-ray transient, discovered by \inte~during
an observation of \GRS. The source presents
strong variations on timescales from seconds to days.
We present results of multiwavelength observations,
including spectral analysis of \inte~observations,
and propose that \targ~is a Galactic X-ray binary.
Further classification of the source is also discussed.

\end{abstract}

\section{Introduction}
X-ray transients are a subclass of X-ray binaries whose 
brightness can vary by a factor of more than 100.
The main reason for those variations are changes in the mass accretion
rate on to the compact object. The variations of the mass accterion rate
can be related to an eccentric orbit, changes in the companion star,
or instabilities of the accretion disk.

\targ~(SIMBAD corrected name IGRJ19140+0951) is an X-ray transient
that was detected by \inte~\citep{diana03a}
on March 6th during an observation of \GRS.
It has also been observed later by Rossi X-ray Timing Explorer (RXTE)
and ground-based optical and radio telescopes such as the 
Nordic Optical Telescope (NOT) and the Giant Meterwave Radio Telescope (GMRT).

\section{X-ray observations}

\inte~observed \GRS~ during revolution 48 (2003 Mar 6-9).
A transient source was discovered \citep{diana03a,diana03b}
in the field, $1^\circ$ from \GRS~ (see Figure 1).

Standard \inte~off-line science analysis software (OSA)
was used for the JEM-X (OSA version 3)
and IBIS/ISGRI (version 2) data reductions.
Mosaic images and combined spectra for the entire \GRS~ pointing
were created following the standard data reduction process
\citep{gol03,wes03}.
The ISGRI 20-40 keV image of this observation is
shown in Figure 1.  The \inte~position of \targ~is
$19^h 13^m 55^s +09^\circ 51' 58''$, with an accuracy of 1'.

A 2.2 ksec RXTE observation was made on 2003 Mar 10 (see also
\citet{swank03}). The RXTE data  was analyzed with the
{\tt rex} script available at the HEASARC web site. Proportional
Counter Array (PCA) spectra
and lightcurves (in 2-8 and 8-20 keV bands) were extracted with
standard screening criteria. The PCA lightcurves (Figure 2) show
some variations on timescales of one hundred seconds, but no
periodicity is found. The observed hardness variations
are not statistically significant. The timescale of the variations
suggests a Galactic origin.

\begin{figure}
\centering
\includegraphics[width=8.0cm]{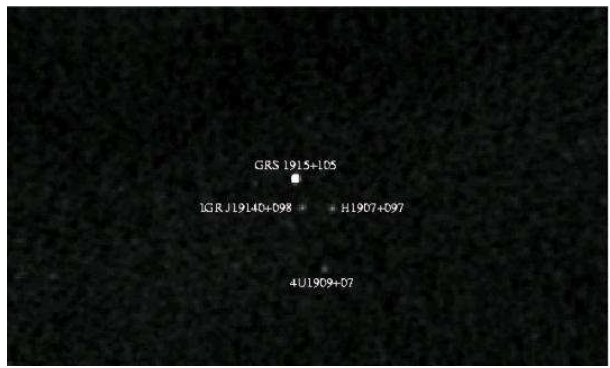}
\caption{
An ISGRI 20-40 keV mosaic image of the \GRS~ field.
For revolution 48, \targ~is clearly detected.
The height is $\sim 12^\circ$ and width $22^\circ$.
North is up and East is to the left.
Adapted from \citet{diana03b}.
\label{fig:ISGRI_image}}
\end{figure}

\begin{figure}
\centering
\psfig{figure=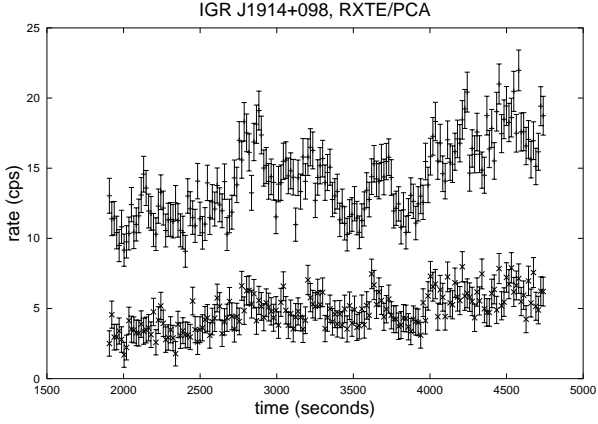,angle=270,width=8.0cm}
\caption{RXTE/PCA lightcurves for March 10th observation, in two
energy bands (upper 2-8 keV, lower 8-20 keV).
Variability on time scales of $\sim 100$ seconds is evident.
No statistically significant periodicities or color variations
were seen during the observation. 
\label{fig:XTE_LC}}
\end{figure}

\begin{figure}
\centering
\psfig{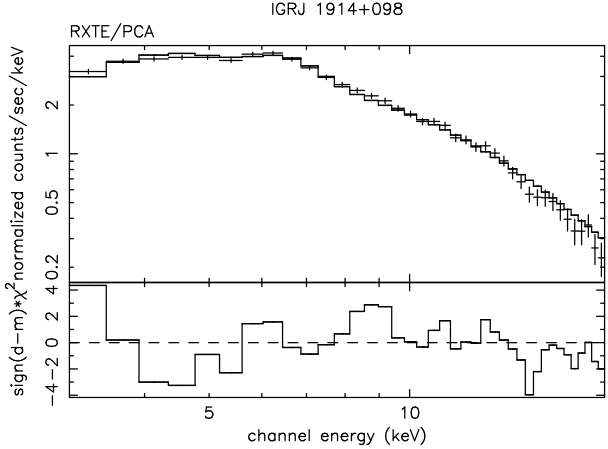}
\caption{The RXTE/PCA spectrum taken on March 10th
\label{fig:XTE_SPEC}}
\end{figure}

\begin{figure}
\centering
\psfig{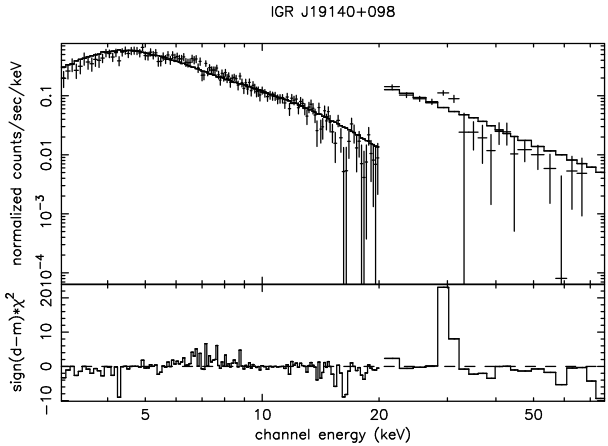}
\caption{The JEM-X and ISGRI spectrum of Revolution 48. 
\label{fig:INT_SPEC}}
\end{figure}

\begin{table}
  \begin{center}
    \caption{The results of the spectral fits. The X-ray flux is in
    the 5-20 keV band and corrected for absorbtion. For details,
    see the text.}\vspace{1em}
    \begin{tabular}[h]{lccc}
      \hline
      & Mar 6-8 & Mar 10 & unit\\
      \hline
      $F_X$ & 38 & 8.5 & $10^{-11} 
      \mathrm{erg} \, \mathrm{cm}^{-2}  \, \mathrm{s}^{-2}$ \\    
       PL $\alpha$ & $2.8 \pm 0.1$ & $1.6 \pm 0.1$ \\
       $N_H$     & 6 (fixed) & $6 \pm 1$ & $10^{22} \, \mathrm{cm}^{-2} $\\    
       $E_C$     & :   & $6.5 \pm 0.2$ & keV \\    
        EW       & :   & $0.6 \pm 0.2$ & keV \\    
       $\chi^2$  & 243 & 54& \\
       dof       & 152 & 40 & \\
       Band      & 3-80 & 3-20 & keV \\
      \hline \\
      \end{tabular}
    \label{tab:fits}
  \end{center}
\end{table}

The spectra  were analyzed with the
XSPEC (version 11.3) software. For PCA, the 3-20 keV energy range was
used, and systematic errors of 1\% have been used in the fitting. 
In the initial absorbed powerlaw fit, 
Large residuals clustering around 6 keV are found when the spetrum
is fitted with an absorbed powerlaw.  
Adding a gaussian line at 6.5 keV to the model removed these residuals.
With this model we obtain a $\chi^2_\nu$  of 1.4. The hydrogen column is 
$6 \pm 1 \times 10^{22} \, \mathrm{cm}^{-2}$. The galactic column in
this direction  is $1.7 \times 10^{22} \, \mathrm{cm}^{-2}$ ({\tt 
ftool nh}), the excess is probably either intrinsic to the system
or caused by small-scale structures of the ISM.
The powerlaw photon index is $1.62 \pm 0.10$, and the
$5-20 \, \mathrm{keV}$ unabsorbed flux is $8.5 \times 10^{-11}
\, \mathrm{erg} \, \mathrm{cm}^{-2} \, \mathrm{s}^{-1}$, corresponding
to $1.0 \times r^2_{\mathrm{kpc}}\times 10^{34}
\, \mathrm{erg} \, \mathrm{s}^{-1}$. The gaussian line
center is at 6.5 keV and equivalent width is 600 eV,
which is very strong for an X-ray binary.

We fitted the JEM-X data between 3-20 keV and ISGRI data between 20-80
keV. Systematic errors of 2\% (10\% for 4-7 keV data
and  20\% for 3-4 keV data) have been added to the JEM-X spectrum.
The data can be fitted with a simple absorbed powerlaw model, with
column density fixed to $6 \times 10^{22} \, \mathrm{cm}^{-2}$.
No normalization constant has been used between the instruments,
as fitting gave a result of $1.00 \pm 0.01$ for the constant.
The fit has $\chi^2_\nu = 1.6$ that could be improved by 0.1 by adding
an iron line, or allowing a freely varying absorbing column. However,
removing the instrumental feature at 30 keV would improve the
$\chi^2_\nu$ by 0.2, so any smaller improvements to the fit can also
be attributed to systematic effects. The powerlaw photon index is
$2.8 \pm 0.1$, and unabsorbed $5-20 \, \mathrm{keV}$ flux during
the INTEGRAL observation is
$3.8 \times 10^{-10} \, \mathrm{erg} \, \mathrm{cm}^{-2} \,
\mathrm{s}^{-1}$, giving a luminosity of
$4.5 \times r^2_{\mathrm{kpc}}\times 10^{34} \,
\mathrm{erg} \, \mathrm{s}^{-1}$.

The powerlaw photon indices of the ISGRI/JEM-X and
PCA spectra are 2.8 vs 1.6, each with an error of 0.1.
The 5-20 keV flux during the \inte~ observation is about four times
higher than during the RXTE observation.
(The 5-20  keV band is used instead of the more common 2-10 as the
former is entirely covered by the instruments and less affected
by absorption or possible low-energy Comptonization cut-off.)
This suggests that a state transition between a 'steep power-law' and
a 'low-hard' state has been observed. Such states are 
characteristic of black hole low-mass X-ray binaries
(see e.g. \citet{mcc03}), and therefore we suggest that
\targ~is a low-mass X-ray binary, possibly containing a black hole.
To determine the nature of donor and compact object,
further observations are needed.

\targ~is in the error circle of the X-ray source EXO~1912+097
\citep{lu96}, so it probably has been detected by EXOSAT.
(See also \citet{intzand04} for BeppoSAX Wide Field Camera
observations of the field.)

\section{Infrared and radio observations}

\begin{figure}
\centering
\psfig{figure=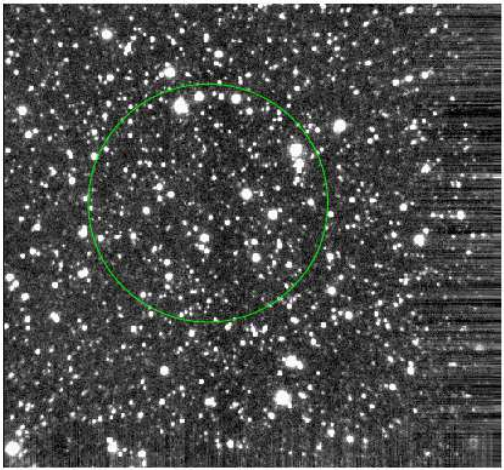,width=8.0cm}
\caption{The field of \targ~imaged in the K-band by NOT, with 
the \inte~1' error circle overlaid.
\label{fig:NOT_K}}
\end{figure}

The \inte~field of \targ~was observed with the Nordic Optical
Telescope (NOT) in spring 2003. Infrared images (JHK) of the field were
taken, but no obvious counterpart for the source could be detected.
Unfortunately, the observations are too short to allow reliable
photometry in any of the used bands. This summer, two nights of NOT
observing time is allocated for further obervations of this field.

The GMRT observations \citep{pandey04} 
revealed one 3.5 mJy source within the error
circle, but a probability of a spurious source
at that brightness level is above 90\%.

\section{Conclusions}

A new X-ray source, \targ~has been discovered with \inte.
Analysis of \inte~and RXTE data shows that the  
spectral and temporal variability of \targ~are
best explained by a Galactic X-ray binary, preferably
one with a low-mass donor and a black hole.
However, further observations are needed to determine
the binary component types.

\section*{Acknowledgments}

J. Schultz acknowledges the financial support of the V\"ais\"al\"a
foundation. D. Hannikainen is a Research Fellow of the Academy of
Finland. O. Vilhu and J. Schultz acknowledge the financial 
support of the Finnish Space technology programme ANTARES, funded by
National Technology Agency TEKES and Academy of Finland.
J. Rodriguez acknowledges financial support
from the French Space Agency (CNES).

We thank Amanda Kaas and John Telting for doing the
NOT TOO observations, and Mamta Pandey and
Jean in't Zand for useful discusssions. 

Based on observations with INTEGRAL, an ESA project with instruments
and science data centre funded by ESA member states (especially
the PI countries: Denmark, France, Germany, Italy, Switzerland,
Spain), Czech Republic and Poland, and with the participation
of Russia and the USA. Based on observations made with
the Nordic Optical Telescope, operated on the island of
La Palma jointly by Denmark, Finland, Iceland, Norway, and Sweden,
in the Spanish Observatorio del  Roque de los Muchachos
of the Instituto de Astrofisica de Canarias. 
This research has made use of data obtained
from the High Energy Astrophysics Science Archive Research Center
(HEASARC), provided by NASA's Goddard Space Flight Center, and
the SIMBAD database, operated at CDS, Strasbourg, France.

\end{document}